# The neural signature of inner peace: morphometric differences between high and low accepters

Alessandro Grecucci[1,2], Parisa Ahmadi Ghomroudi[1], Bianca Monachesi[1], Irene Messina[3]

[1]DiPSCo – Department of Psychology and Cognitive Sciences, University of Trento, Rovereto, Italy

[2]CISMed – Center for Medical Sciences, University of Trento, Trento, Italy

[3] University of Mercatorum, Rome, Italy

**Corresponding author:**

Parisa Ahmadi Ghomroudi

Department of Psychology and Cognitive Sciences,

University of Trento,

Corso Bettini, 84, 38068,

Rovereto, Italy

**E-mail**: p.ahmadighomroudi@unitn.it

**Tel:** (+39) 0464 808302

**Fax:** (+39) 0464 808602

**Abstract**

Acceptance is an adaptive emotion regulation strategy characterized by an open and non-judgmental attitude toward mental and sensory experiences. While a few studies have investigated the neural correlates of acceptance in task-based fMRI studies, a gap remains in the scientific literature in dispositional use of acceptance, and how this is sedimented at a structural level. Therefore, the aim of the present study is to investigate the neural and psychological differences between infrequent acceptance users (i.e., low accepters) and frequent users (i.e., high accepters). Another question is whether high and low accepters differ in personality traits and emotional intelligence. To this aim, we applied, for the first time, a data fusion unsupervised machine learning approach (mCCA-jICA) to the gray matter (GM) and white matter (WM) of high accepters (N = 50), and low accepters (N = 78) to possibly find joint GM-WM differences in both modalities. Our results show that two covarying GM-WM networks separate high from low accepters. The first network showed decreased GM-WM concentration in a fronto-temporal-parietal circuit largely overlapping with the Default Mode Network, while the second network showed increased GM-WM concentration in portions of the orbito-frontal, temporal, and parietal areas, related to a Central Executive Network. At the psychological level, the high accepters display higher openness to experience compared to low accepters. Overall, our findings suggest that high accepters compared to low accepters differ in neural and psychological mechanisms. These findings confirm and extend previous studies on the relevance of acceptance as a strategy associated with well-being.

## Introduction

The ability to regulate emotion is believed to be a core competence for mental health and well-being. Difficulties in regulating emotions have been associated with a broad spectrum of psychological disorders (Kring & Sloan, 2009; Sheppes et al., 2015; Dadomo et al., 2018; Frederickson et al., 2018). Adaptive emotion regulation strategies are considered effective for individuals experiencing emotion dysregulation and are commonly incorporated into various treatment (Leahy et al., 2011; Dadomo et al., 2016, 2018; Grecucci et al., 2017; Panfilis, 2019). Additionally, acceptance is believed to be a core construct in third wave behavioral therapies (Hayes, 2004; Kahl et al., 2011) and experiential-dynamic approaches (Frederickson et al., 2018; Grecucci et al., 2020; Messina et al., 2020). In these contexts, acceptance is defined as "*the active and ware embrace of private experiences without unnecessary attempts to change their frequency or form*"(Hayes et al., 2012). This study views acceptance as the counter of experiential avoidance, as it involves a mental stance of open curiosity towards ongoing mental and sensory experiences (Grecucci et al., 2015; Goldin et al., 2019). Acceptance is considered an adaptive strategy and is positively associated with wellbeing (Aldao et al., 2010). In one study that examined different adaptive and maladaptive emotion regulation strategies across eight situations, Aldao and Nolen-Hoeksema (2012) found that acceptance was a particularly adaptive strategy. Another study by Dan-Glauser and Gross (2015) compared acceptance to no regulation and found that acceptance led to an increase in positive emotions and emotional expressivity, as well as a decrease in respiration rate. Goldin (2019) and colleagues also reported that acceptance resulted in a decreased respiration rate and negative emotion, with no significant difference in skin conductance level compared to no regulation.

**Neural correlates of acceptance**

Previous functional neuroimaging studies have investigated the neural correlates of acceptance through task-based fMRI. Kross and colleagues (2009) reported reduction in activity in the anterior cingulate cortex and medial prefrontal cortex during regulation of negative autobiographical memories by acceptance, compared to no regulation. Another fMRI study by Smoski and colleagues (2014) showed increased BOLD activity in dorsomedial, anterior midcingulate, left lateral prefrontal cortex (PFC), right dorsolateral PFC regions during acceptance vs no regulation while viewing sad images. Goldin and colleagues (2019) reported greater BOLD activity in attention and control regions such as the medial prefrontal cortex, dorsolateral prefrontal cortex, ventrolateral prefrontal cortex, during reading of negative autobiographical scripts in acceptance vs no regulation. A fMRI study conducted by Dixon and colleagues (2020) depicted involvement of frontoparietal regions including rostrolateral PFC, inferior frontal sulcus, middle frontal gyrus, anterior midcingulate cortex, and posterior middle temporal gyrus during acceptance vs no regulation while reacting to negative self-belief.

To provide a coherence to the sparse findings of the literature on acceptance, Messina and colleagues (2021) conducted a meta-analysis of 13 fMRI studies. Results showed increased in executive areas (although only in a subgroup of studies), and more importantly, a major cluster of decreased brain activity located in the posterior cingulate cortex (PCC)/precuneus. These findings casted a shadow on a key tenet of neuroscientific models of emotion regulation underlining the necessary involvement of high-level executive processes.

A recent meta-analysis of 42 fMRI studies comparing reappraisal and acceptance confirmed and extended previous results (Monachesi et al., 2023). The results revealed that reappraisal was associated with increased activity in the superior frontal gyrus and left middle frontal gyrus, as well as decreased activity in the left globus pallidus and putamen, whereas acceptance was associated with reduced activity in the posterior cingulate cortex (PCC)/precuneus, a key region of the default mode network (DMN). Moreover, acceptance was associated with increased activity in the ventrolateral prefrontal cortex (VLPFC) and claustrum (Monachesi et al., 2023). Several studies have reported decreased brain activity within default mode network (DMN) regions during acceptance (Opialla et al., 2015; Dixon et al., 2020). The DMN, associated with mind-wandering (Christoff et al., 2009),has been perceived as the opposite of mindfulness(Mrazek et al., 2012). Such findings imply the disruption of ruminative and self-reflective processes independent of executive processes (Ellard et al., 2017; Messina et al., 2021). One intriguing possibility is that such reduction in the activity of the DMN is also associated with a reduction at a structural level (gray matter concentration) in regions ascribed to the DMN. Recent studies have shown that resting state macro-networks can also be found at a structural level using Independent Component Analysis (ICA) based methods(Meier et al., 2016; Vanasse et al., 2021; Baggio et al., 2023; Grecucci et al., 2023).

**The link between acceptance and personality traits**

Personality traits refer to an individual’s relatively stable patterns of thoughts, feelings, and actions (McCrae & Costa Jr., 2003). The Five-Factor Model (FFM) of Personality is widely recognized as the most influential and prominent model in psychology (Costa and McCrae, 1989; Costa and McCrae, 1992). According to this model, there are five primary dimensions in which individuals differ in their personality traits: Neuroticism (N): Reflects an individual's tendency to experience negative emotions and psychological distress in response to stressful situations. Extraversion (E): Measures an individual's level of sociability, positive emotionality, and overall activity. Openness to Experience (O): Assesses an individual's curiosity, independent judgment, and willingness to embrace new experiences while considering levels of conservatism. Agreeableness (A): Evaluates an individual's tendencies toward altruism, sympathy, and cooperation in interpersonal relationships. Conscientiousness (C): Assesses the extent to which an individual exhibits self-control in planning, organization, and overall accountability. Several studies have explored the relationship between personality traits and emotion regulation strategies. Barańczuk, (2019) explored the relationship between the five-factor model of personality and adaptive as well as maladaptive emotion regulation strategies. The results revealed that higher levels of extraversion, openness to experience, agreeableness, and conscientiousness were positively correlated with the use of adaptive strategies, such as reappraisal, problem-solving, and mindfulness, whereas lower levels of neuroticism were negatively correlated with the use of these adaptive strategies. Conversely, higher levels of neuroticism were positively correlated with maladaptive strategies like avoidance, suppression, worry, and rumination, while lower levels of extraversion, openness to experience, agreeableness, and conscientiousness were negatively correlated with the use of these maladaptive strategies. Additionally, openness to experience showed a positive correlation with reappraisal, problem solving, mindfulness and it was inversely associated with suppression. A meta-analysis of 141 studies examined the relationships between trait mindfulness and personality traits. The findings revealed negative correlations with neuroticism, positive correlations with conscientiousness, and openness to experience (Banfi & Randall, 2022). As far as we know, no study has reported differences between high and low accepters in terms of the personality factors.

**The link between acceptance and emotional intelligence**

Trait Emotional Intelligence (tEI) refers to the ability to comprehend and use emotions intrapersonally and interpersonally. It is defined as “a constellation of emotional self-perceptions located at the lower levels of personality hierarchies” (Petrides, Pita, Kokkinaki, 2007). Some studies have revealed that individuals with high trait emotional intelligence exhibit greater non-judgmental attention to the present moment (Charoensukmongkol, 2014; Wang and Kong, 2014) and engage in less rumination compared to those with low trait emotional intelligence (Ramos et al., 2007; Salguero et al., 2013). A study conducted by Zanella and colleagues (2022) investigating the connection between tEI and a few Emotion Regulation strategies revealed a negative relationship between tEI and certain maladaptive emotion regulation strategies, specifically suppression and self-blame. Notably, no significant association was observed between tEI and acceptance. However, as far as we know, no study clearly explored the differences in terms of tEI subscales between high and low accepters.

**The present study**

Previous studies have explored the regulatory nature of acceptance when participants applied this strategy to emotional stimuli during task-based fMRI. However, there is a dearth of research addressing how the dispositional use of emotion regulation strategies are encoded in the brain, how individuals with varying degrees of acceptance ability are sedimented at a structural level, and the potential contributions of white matter, which have been largely overlooked(Giuliani et al., 2011; Kühn et al., 2011; Pappaianni et al., 2018; Ahmadi Ghomroudi et al., 2023). Furthermore, although some studies have shown gray matter differences related to other emotion regulation strategies(Giuliani, Drabant, & Gross, 2011; Pappaianni et al., 2020) the present study aims to investigate the differences in both gray and white matter in the brains of individuals with high and low levels of acceptance for the first time. Additionally, we will explore eventual differences in personality traits and emotional intelligence between individuals with high vs low acceptance, in order to overcome the limitations in the existing literature.

The study has a threefold aim: firstly, to identify independent circuits of covarying gray and white matter that differ between individuals with high versus low levels of acceptance, using a data fusion machine learning method. We expect to observe differences in gray and white matter concentration in regions of the default mode network (DMN), particularly the PCC/precuneus which has been linked to reduced mind wandering and rumination among high accepters (Opialla et al., 2015; Dixon et al., 2020; Messina et al., 2021; Monachesi et al., 2023) . Alternatively, increased gray and white matter concentration in prefrontal regions involved in inhibiting emotional regions (Morawetz et al., 2017) may be observed. Based on previous studies, we hypothesize that subcortical regions (e.g., insula, cingulate gyrus cortex) and frontal regions (e.g., ventrolateral prefrontal cortex) may exhibit increased gray and white matter concentration for high accepters, whereas regions like PCC/precuneus may display reduced concentration in high accepters (Messina et al., 2021; Monachesi et al., 2023). Secondly, it aims to identify possible differences between high and low accepters in terms of personality traits. Drawing from a previous study (Barańczuk, 2019), we hypothesize that high acceptance is positively associated with traits such as openness to experiment, conscientiousness and negatively correlated with traits such as neuroticism Finally, it aims to explore potential differences in emotional intelligence (EI) between high and low accepters. Building on previous studies, we hypothesize that high acceptance is positively associated with emotional intelligence (Chu, 2010; Schutte and Malouff, 2011). One possibility is that high accepters differ from low accepters in several dimensions associated with personality traits and emotional intelligence. Alternatively, the link between high acceptance and emotional intelligence may be more restricted to specific dimensions.

Of note, from a methodological point of view, our study departs from the previous attempts to characterize emotion regulation abilities who largely relied on univariate methods, by using a multivariate unsupervised machine learning approach. In fact, the present study uses multimodal canonical correlation analysis (mCCA) in combination with joint independent component analysis (jICA), (Sui et al., 2012), a multivariate machine learning technique. Previous studies have applied univariate statistical techniques, such as voxel-based morphometry, which do not account for voxel dependencies across the entire brain (Mak et al., 2009; Giuliani, Drabant, Bhatnagar, et al., 2011; Hermann et al., 2013) .Conversely, multivariate approaches such as Independent Component Analysis (ICA) based methods (Norman et al., 2006; Xu et al., 2009; Sorella et al., 2019; Grecucci et al., 2023) and mCCA+jICA, consider the statistical dependencies among voxels and can identify complex and sparse patterns throughout the entire brain. As a result, these approaches do not require defining a priori regions of interest (ROIs) that are typically based on brain atlases derived from histological properties and biologically implausible parcellations of the brain. The mCCA+jICA merges two modalities, such as GM and WM, to identify the correlations between them using mCCA and separate the covariance matrix into independent networks of covarying GM-WM using jICA. The fusion of these two methods allows for a multimodal fusion (MMF) approach to identify the unique and shared variance associated with each modality in relation to cognitive functioning in individuals. This method has been successfully implemented in several studies. For example,  Liang and colleagues (2021) applied mCCA-jICA to investigate the covariation of gray and white matter concentration in cognitive decline and mild cognitive impairment, Baggio and colleagues (2023) used this method to investigate individual differences in anxiety trait, and Kim and colleagues (2015) employed mCCA-jICA to explore gray and white matter networks that contribute to structural alterations in the brains of patients with OCD. The multi-modal fusion approach provides a reliable interrelationship of changes in each modality, yielding more integrated information about the brain. Multi-modal fusion approaches are particularly suitable for analyzing complicated and weak effects that may be undetectable in high-dimensional data sets, while also exhibiting robustness to noise (Calhoun & Sui, 2016).

## Materials & Methods

### Participants

T1 structural brain images, CERQ, TEIQue-SF questionnaire scores, and NEO-Five-Factor Inventory (NEO-FFI) of one group of high accepters (n = 58, 16 Females, mean acceptance score 9.68 ±1.21,mean age 29,48 ± 12,2 and average 12.5 ± 0.91 years of education), and another group of low accepters (n = 70, 20 Females, mean acceptance score 4.84 ± 1.71, mean age 29,92 ± 12,7 years and average 12.5 ± 0.91 years of education) of German native speakers were included in this study. The data was selected from “Leipzig study for mind-body-emotion interactions” (OpenNeuro database, accession number ds000221) (LEMON). The data collection was conducted at the Max Planck Institute for Human Cognitive and Brain Sciences (MPI CBS) in Leipzig (Babayan et al., 2019) The exclusion criteria for data collection were as follows: no cardiovascular disease, history of psychiatric diseases, history of neurological disorders, history of malignant diseases or intake of centrally active medication, beta- and alpha-blocker, cortisol, any chemotherapeutic or psychopharmacological medication. We also excluded participants who reported drug or excessive alcohol use.

A total of 128 participants (36 females) were included in the study after excluding 7 participants due to corrupted data. Participants in the study provided written informed consent prior to the data collection and agreed to the anonymous data sharing. They received compensation for participating in the study after the completion of all assessments.

### Behavioral Data

The study used three questionnaires to assess various constructs. The first questionnaire utilized was the Cognitive Emotion Regulation Questionnaire (CERQ) German version (Garnefski et al., 2001; Loch et al., 2011). The CERQ measures nine cognitive coping strategies, including self-blame, acceptance, rumination, positive refocusing, refocus on planning, positive reappraisal, putting into perspective, catastrophizing, and blaming others. The German version of the questionnaire consists of 27 items, with three questions measuring each strategy on a 5-point Likert-type scale ranging from 1 (almost never) to 5 (almost always). For the purposes of this study, the Acceptance scale was specifically focused on to differentiate the two groups and investigate neural and psychological differences between high and low accepters. The second questionnaire employed was the German adaptation of the NEO-Five-Factor Inventory (Borkenau & Ostendorf, 2008) to evaluate the Big Five of Personality Inventory (NEO-FFI; Costa & McCrae, 1992). This questionnaire consists of 60 items, which are categorized into five personality factors: Neuroticism, Extraversion, Openness to experience, Agreeableness, and Conscientiousness. Responses are provided on a 5-point Likert scale, ranging from 0 (strong denial) to 4 (strong approval). Lastly, the German adaptation (TEIQue-SF, Freudenthaler et al., 2008) of the Trait Emotional Intelligence Questionnaire-Short Form (TEIQue-SF; Petrides & Furnham, 2004) was used. This questionnaire comprises four subscales: Well-Being, Self-Control, Emotionality, and Sociability. It includes a total of 30 items, with two items from each of the 15 facets of the TEIQue. To investigate whether the frequency of using acceptance leave a structural trace in the brain, participants were divided into two subgroups based on the median value (-0.0136) of the z scores of the Acceptance subscale of the CERQ questionnaire. Individuals with scores below the median were assigned to low acceptors group (70 individuals) and those with scores above the median were assigned to high accepters' group (58 individuals).

**Image Acquisition**

The T1-weighted structural images were acquired using a 3 Tesla scanner (MAGNETOM Verio, Siemens Healthcare GmbH, Erlangen, Germany) equipped with a 32-channel head coil. The scanner was not subject to any major maintenance or updates during the experiment and the data quality remained stable throughout the study.

## Data analysis

### *Pre-processing*

First, structural MRI scans were initially inspected manually to assess their quality, ensuring the exclusion of any possible artifacts. The structural images were then reoriented manually to the anterior commissure as the reference point. Next the data was pre-processed using Computational Anatomy Toolbox (CAT12, http://www.Neuro.uni-jena.de/cat/), a toolbox for statistical Parametric Mapping (SPM12) in MATLAB environment (The Mathworks, Natick, MA, USA).

The CAT12 toolbox was used to segment the Images into gray matter (GM), white matter (WM), and cerebrospinal fluid (CSF). To enhance registration accuracy, Diffeomorphic Anatomical Registration of GM images was conducted using Exponential Lie algebra (DARTEL) tools for SPM12 instead of traditional whole brain registration (Yassa & Stark, 2009; Grecucci et al., 2016; Pappaianni et al., 2018). Finally, the DARTEL images were normalized to the MNI-152 standard space and smoothed using a 12-mm full-width at half-maximum (FWHM) Gaussian kernel [12, 12, 12].

### *Data fusion unsupervised machine learning method*

The fusion analysis of mCCA-jICA was performed using the Fusion ICA Toolbox (FITv2.0d, http://mialab.mrn.org/software/fit/) (Figure 1). First, GM and WM features were extracted, which resulted in two feature matrices with rows representing the number of participants and columns representing the number of voxels. These matrices were then normalized to z-scores. After applying the default parameters, a total of 12 independent components were extracted. The number of components was estimated for both modalities using the minimum description length criterion (Li et al., 2007), Finally, singular value decomposition (SVD) was used to reduce dimensionality of feature matrices.

- - - -

Please insert Figure 1

- - - -

**Backward logistic regression Analysis on neural data**

To find which IC correctly classifies high and low accepters backward logistic regression was performed in (JASP Team, 2021) backward regression analysis is a widely used wrapper method for feature selection. It involves iteratively removing the least statistically significant independent variable from an initial model containing all independent variables, until a model with optimal performance is obtained (Jollans et al., 2019). Backward logistic regression is a specific type of backward regression that is commonly used when the dependent variable is binary. Backward regression is especially useful when the number of independent variables is not large Smith (2018). In this study high and low accepter scores were entered as dependent variable and age, gender and twelve loading coefficients were entered in as independent variables.

**Behavioral Analysis**

Statistical analysis of behavioral data was conducted using JASP Team (2021). JASP (Version 0.16). Welch's t-test was conducted in order to find difference in high and low accepters in four subscales of five personality factors: Neuroticism, Extraversion, Openness to experience, Agreeableness, and Conscientiousness and emotional intelligence: Well-Being, Self-Control, Emotionality, and Sociability and. To control for Type I error, a false discovery rate (FDR) correction was applied to the p-values

## Results

### *Covarying GM-WM networks that separate high from low accepters*

mCAA and jICA was applied to structural data of 128 individuals and returned 12 independent circuits of covarying gray and white matter components. The number of components was estimated by the software according to the Information theoretic criteria. The values obtained from these networks reflect changes in gray and white matter concentration. Positive values indicate an increase in concentration, while negative values indicate a decrease. Covariation between gray and white matter components suggests a similar pattern of concentration changes in both types of matter. See Figure 2.

-------------------------------------------------

Please insert figure 2 about here.

-------------------------------------------------

### *Backward logistic regression Analysis*

The Backward logistic regression to characterize high vs low accepters returned as significant the IC7 ($\beta$ = -9.157, SE = 4.143, Wald = 5.357, P = 0.027), which showed decreases GM-WM concentration in fronto-temporal-parietal regions, and IC10 ($\beta$ = 21.717, SE = 9.173, Wald = 5.756, P= 0.018), which showed increased GM-WM concentration in the parts of orbito-frontal, temporal and parietal cortices. See Table 1, 2.

-------------------------------------------------

Please insert figure 3 about here.

-------------------------------------------------

-------------------------------------------------

Please insert figure 4 about here.

-------------------------------------------------

-------------------------------------------------

Please insert table 1 about here.

-------------------------------------------------

-------------------------------------------------

Please insert table 2 about here.

-------------------------------------------------

***Behavioral results***

The Welch's t-test was conducted to examine the difference between low and high accepters in terms of emotional intelligence and five personality factors. The result revealed significant difference between low and high accepters, with FFM of Personality - openness to experience higher in high accepters $t(122.92) = -1.760$, $p = 0.040$, $pFDR = 0.040$). See Table 3.

-------------------------------------------------

Please insert table 3 about here.

-------------------------------------------------

-------------------------------------------------

Please insert figure 5 about here.

-------------------------------------------------

-------------------------------------------------

Please insert figure 6 about here.

-------------------------------------------------

**Discussion**

The aim of this study was to find joint gray and white matter differences that characterize high accepters compared to low accepters. To this aim a novel data fusion unsupervised machine learning approach known as mCCA+jICA was used. mCCA+jICA detected two brain networks which significantly differentiate high accepters and low accepters. The first network included co-altered gray and white matter concentration in the default system (DS), in the Salience Network (SN) and in subcortical areas (basal ganglia and hippocampus), with reduced matter concentration for high accepters. The second network included co-altered gray and white matter concentrations in several executive areas in the prefrontal (dorsomedial prefrontal and orbito-frontal cortices) and parietal cortex, with higher gray matter concentration for high accepters. In what follows, we discuss more in details the role of these networks.

**Network 1: Reduced Fronto-temporal-parietal network**

Network 1 included several areas of the default system (posterior and anterior midline structures, anterior temporal areas, angular gyrus), the dorsal anterior cingulate cortex (dACC) and right insula, and a series of subcortical structures, resulted joint, co-altered components that differ between high and low accepters, with lower matter concentration in high accepters. The implication of the default system in acceptance is in line with previous functional MRI studies which have reported decreased activity in the default system and decreased connectivity within the default system areas. These changes have been associated with individual differences in acceptance-based meditation experience (Brewer et al., 2011) and with acceptance considered as a component of trait mindfulness (Wang et al., 2014; Harrison et al., 2019). In task-related studies of acceptance, it has been suggested that reduced DMN activity, especially in the posterior cingulate and adjacent areas, may reflect the ‘interruption’ of ruminative processing that characterizes acceptance (Ellard et al., 2017; Messina et al., 2021). On the contrary, hyper-activation of the DMN characterizes rumination and related dysregulated states (Sheline et al., 2009; Messina et al., 2016a; Whitfield-Gabrieli and Ford, 2012; Buckner et al., 2019). According to semantic models of emotion regulation (Viviani, 2013; Messina et al., 2015, 2020), the DMN may contribute to emotion regulation by modulating the semantic processing of emotional stimuli, with automatic activation of internal semantic representations corresponding to higher DMN activation. In line with such model, decreased gray and white matter concentration in association to higher use of acceptance may correspond to a more detached attitude toward emotionally relevant semantic representations, contrasting with a more engaged attitude toward internally generated contents

The concept of a detached attitude toward internally generated semantic representations aligns with the co-alteration observed in the basal ganglia (lentiform nucleus) and in the hippocampus in the present study, as well as in a previous study on trait mindfulness (Taren et al., 2013). As part of the basal ganglia, the role of lentiform nucleus in the reward response has been well established (Schultz, 2016). And, the joint activation of basal ganglia and hippocampus structures seems to contribute to reward mechanisms by forming and storing memories of events and places of rewarding experiences (Singer & Frank, 2009; Lo Gates et al., 2018). The finding of reduced gray matter in reward-related subcortical regions in high accepters is in line with previous findings that reported less susceptibility to extrinsic incentive in expert meditators compared to non-meditators (Kirk et al., 2015; Brown et al., 2013), reflecting a sort of 'inner peace' state in meditators. Also, the right insula and the anterior cingulate cortex (dACC) may contribute to the achievement of this 'let it be' attitude toward internal arousing representations. These areas are key nodes of the salience network, which integrates internal and external information to guide behavior (Seeley, 2019). The salience network is responsible for switching between the DMN and the central executive network, signalling the DMN to reduce its activity when attention is shifted from internal to external focus (Goulden et al., 2014; Jilka et al., 2014). Taken together, these results account for a difference between high accepters and low accepters in the management of internally generated arousing contents.

**Network 2: Orbito-frontal, temporal and parietal**

The second network detected in the present study was composed by a series of prefrontal and superior parietal areas, with a possible overlap with the dorsal fronto-parietal network associated with top-down attention processes (Cole et al., 2013, 2014b; Dodds et al., 2011; Power et al., 2011; Scolari et al., 2015). The activation of this fronto-parietal executive network (together with the insula and supplemental motor areas, also detected as part of this network) has been largely described as associated with voluntary emotion regulation (for meta-analyses see Messina et al., 2015; Morawetz et al., 2017; Monachesi et al., 2023). Thus, our findings suggest that high accepters could be more able in recruiting cognitive and attention regulatory brain networks to down-regulate emotion.

Among prefrontal areas, the dorsomedial prefrontal cortex (with extensions to the ACC and SMA) has deserved particular attention in the context of the research on mindfulness because it is involved in meditative acceptance-based processes (for a meta-analysis see Tomasino et al., 2013). Moreover, the DMPFC has been previously reported as more activated during emotion regulation tasks in individuals with higher trait mindfulness (Frewen et al., 2010; Modinos et al., 2010). Aside from emotion regulation, the DMPFC has been associated with evaluation of self-referential stimuli (Gusnard et al., 2001; Northoff & Bermpohl, 2004). For this reason, activity in DMPFC regions are thought to support meta-cognitive reflective awareness of internal states (Olsson & Ochsner, 2008; Lutz et al., 2016). Consistently, the higher gray matter concentration in high accepters observed in the present study may reflect the ability of being more aware of internal states.

**Personality traits, Emotional intelligence, and acceptance**

The result of this study revealed greater level of openness to experience among high accepters. Individuals with a high openness to experience score tend to exhibit curiosity, imagination, open-mindedness, and a propensity to embrace unconventional ideas (Barrick et al., 2001). These individuals show a keen interest in both their inner thoughts and the external world (Costa & McCrae, 1992). The commonality observed between individuals with a high acceptance ability and openness to experience is their propensity for maintaining an open and curious mindset, continuously engaged with the flow of experiences (Giluk, 2009; Barańczuk, 2019; Banfi & Randall, 2022) High acceptance ability might play a role in facilitating and supporting the willingness to embrace new experiences associated with openness to experience, helping individuals maintain curiosity towards ongoing experiences. The results of this study show no significant relationship between high and low accepters and emotional intelligence, aligning with prior research findings (Mikolajczak et al., 2008; Zanella et al., 2022). However, this result contradicts our initial hypothesis, suggesting the necessity for in-depth further investigation and exploration

**Conclusions and Limitations**

These findings offer valuable insights into the neural underpinnings low and high accepters, which may have implications for the development of neuromodulation treatments targeting specific brain circuits to enhance acceptance abilities. However, there are some limitations to note. The reliance on self-report questionnaires in this study may have introduced biases due to participants' limited self-awareness. Additionally, the current study exclusively used structural data, and future research would benefit from integrating functional data to further explore these findings. Furthermore, while our study included a large sample size, future studies with even larger samples are necessary to validate our results.

**Statements & Declarations**

**Funding**

This research did not receive any specific grant from funding agencies in the public, commercial, or not-for-profit sectors

**Competing Interests**

The authors have no competing interests to declare.

**Author contributions**

Alessandro Grecucci, conceived the work, analysed data, reviewed the manuscript and supervised the work. Parisa Ahmadi Ghomroudi. analysed data, wrote the manuscript. Bianca Monachesi: reviewed the manuscript. Irene Messina wrote and reviewed the manuscript

**Data availability**

The complete LEMON Data is available through Gesellschaft für wissenschaftliche Datenverarbeitung mbH Göttingen (GWDG) at this URL: https://www.gwdg.de/ . Both raw and preprocessed data can be accessed via a web browser at https://ftp.gwdg.de/pub/misc/MPI-Leipzig_Mind-Brain-Body-LEMON/ or through a high-speed FTP connection at ftp://ftp.gwdg.de/pub/misc/MPI-Leipzig_Mind-Brain-Body-LEMON/. In the event that the data location changes in the future, the location of the dataset can be resolved with PID 21.11101/0000-0007-C379-5 (e.g. http://hdl.handle.net/21.11101/0000-0007-C379-5 )

**Table 1**

**Result from backward logistic regression analysis with winning models IC7 and IC10 for high and low acceptors**

| Source | β | SE | Wald χ2 | p | OR | 95% CI |
|---|---|---|---|---|---|---|
| IC7 | -9.157 | 4.143 | 5.375 | 0.027 | $1.055 \times 10^{-4}$ | [0, 0.354] |
| IC10 | 21.717 | 9.173 | 5.756 | 0.018 | $2.702 \times 10^{+9}$ | [2.775, 22.47] |

*Note:* OR = Odds ratio, CI = Confident interval, SE= Standard Error

**Table 2**

**Gray and White Independent Component 7, 10. Talairach labels of regions of interest, Brodmann area, volume (expressed in cc) and max values coordinates are shown.**

| Network | Area | Brodmann Area | Volume (cc) | Random effects: Max Value (x, y, z) |
|---|---|---|---|---|
| IC7 GM | Cerebellar Tonsil | * | 0.4/3.8 | 4.9 (-34, -38, -35)/8.4 (46, -48, -35) |
| | * | * | 0.1/0.4 | 3.7 (-10, 4, -7)/6.1 (1, 19, 53) |
| | Paracentral Lobule | 4, 5, 6, 31 | 2.4/2.2 | 6.8 (-1, -30, 61)/6.5 (4, -30, 61) |
| | Medial Frontal Gyrus | 6, 8 | 1.6/3.0 | 6.7 (-3, -26, 61)/6.6 (4, -25, 61) |
| | Declive | * | 2.5/0.0 | 6.7 (-18, -82, -17)/-999.0 (0, 0, 0) |
| | Middle Frontal Gyrus | 6, 8, 9, 10 | 3.2/1.9 | 5.3 (-31, 38, 26)/6.0 (25, 34, 34) |
| | Superior Frontal Gyrus | 6, 8, 9, 10 | 1.7/3.4 | 5.9 (-28, 51, -1)/6.0 (13, 9, 59) |
| | Sub-Gyral | 6 | 3.2/3.2 | 5.3 (-42, 23, 18)/6.0 (30, -76, 23) |
| | Lentiform Nucleus | * | 1.0/2.2 | 4.8 (-16, 3, -3)/5.9 (21, 0, -1) |
| | Posterior Cingulate | 18, 29, 30 | 0.7/1.7 | 5.1 (-19, -58, 10)/5.7 (18, -66, 12) |
| | Culmen | * | 0.5/1.1 | 4.1 (-12, -63, -3)/5.7 (46, -48, -30) |
| | Precentral Gyrus | 4 | 0.6/1.7 | 4.5 (-9, -23, 63)/5.6 (21, -23, 59) |
| | Precuneus | 7, 19 | 1.2/0.5 | 5.5 (-25, -59, 42)/4.6 (27, -76, 20) |
| | Extra-Nuclear | * | 0.4/1.1 | 4.6 (-19, -53, 10)/5.4 (22, -55, 8) |
| | Inferior Parietal Lobule | 40 | 0.1/1.3 | 4.2 (-31, -56, 43)/5.3 (53, -41, 27) |
| | Lingual Gyrus | 17, 18, 19 | 2.5/0.8 | 5.3 (-24, -70, -4)/5.0 (15, -51, 5) |
| | Superior Parietal Lobule | 7 | 0.8/0.0 | 5.3 (-24, -60, 46)/-999.0 (0, 0, 0) |
| | Cuneus | 17, 18, 19, 23, 30 | 0.9/1.8 | 4.9 (-12, -71, 10)/5.1 (28, -79, 26) |

| Network | Area | Brodmann Area | Volume (cc) | Random effects: Max Value (x, y, z) |
|---|---|---|---|---|
| | Parahippocampal Gyrus | 30 | 0.1/0.6 | 3.6 (-24, -10, -16)/5.1 (12, -48, 4) |
| | Uvula | * | 0.4/0.0 | 5.0 (-19, -78, -23)/-999.0 (0, 0, 0) |
| | Middle Temporal Gyrus | 21, 38 | 0.3/0.4 | 4.0 (-48, 8, -18)/5.0 (33, -76, 20) |
| | Postcentral Gyrus | 3, 4, 5 | 1.2/0.7 | 4.8 (-21, -32, 60)/5.0 (22, -26, 62) |
| | Fusiform Gyrus | 18, 19 | 0.6/0.0 | 4.8 (-21, -88, -15)/-999.0 (0, 0, 0) |
| | Middle Occipital Gyrus | 18, 19 | 0.0/0.4 | -999.0 (0, 0, 0)/4.8 (30, -79, 21) |
| | Inferior Frontal Gyrus | 46 | 0.3/0.4 | 4.2 (-45, 23, 15)/4.8 (48, 21, 17) |
| | Supramarginal Gyrus | * | 0.1/0.4 | 3.7 (-50, -51, 25)/4.7 (53, -41, 31) |
| | Superior Temporal Gyrus | 13, 22, 38 | 1.5/1.2 | 4.6 (-45, 11, -19)/4.5 (43, 9, -24) |
| | Tuber | * | 0.0/0.2 | -999.0 (0, 0, 0)/4.6 (46, -54, -30) |
| | Superior Occipital Gyrus | * | 0.0/0.1 | -999.0 (0, 0, 0)/4.6 (33, -76, 26) |
| | Cingulate Gyrus | 24, 32 | 0.2/1.0 | 4.2 (-1, -8, 47)/4.4 (4, 25, 36) |
| | Lateral Ventricle | * | 0.0/0.2 | -999.0 (0, 0, 0)/4.2 (28, -55, 8) |
| | Anterior Cingulate | 32 | 0.3/0.0 | 4.0 (-1, 30, 25)/-999.0 (0, 0, 0) |
| | Angular Gyrus | * | 0.1/0.0 | 3.9 (-31, -58, 37)/-999.0 (0, 0, 0) |
| | Inferior Semi-Lunar Lobule | * | 0.0/0.1 | -999.0 (0, 0, 0)/3.8 (25, -69, -36) |
| | Pyramis | * | 0.0/0.1 | -999.0 (0, 0, 0)/3.7 (21, -72, -32) |
| | Inferior Temporal Gyrus | * | 0.0/0.1 | -999.0 (0, 0, 0)/3.6 (59, -15, -16) |
| | Subcallosal Gyrus | 34 | 0.0/0.1 | -999.0 (0, 0, 0)/3.6 (15, 4, -13) |
| | Inferior Occipital Gyrus | 18 | 0.1/0.0 | 3.6 (-21, -88, -9)/-999.0 (0, 0, 0) |
| | | | | |
| IC7 WM | Extra-Nuclear | * | 1.5/1.0 | 6.9 (-19, -37, 21)/5.4 (21, -35, 21) |
| | Precuneus | 7, 19, 39 | 3.1/2.8 | 5.7 (-27, -50, 52)/6.7 (13, -62, 51) |

| Network | Area | Brodmann Area | Volume (cc) | Random effects: Max Value (x, y, z) |
|---|---|---|---|---|
| | Inferior Parietal Lobule | 7, 39, 40 | 4.4/2.4 | 6.5 (-46, -42, 42)/5.4 (55, -35, 27) |
| | Lateral Ventricle | * | 1.0/0.5 | 6.5 (-16, -34, 21)/5.1 (19, -31, 22) |
| | Middle Temporal Gyrus | 21, 39 | 1.6/2.5 | 6.4 (-49, 2, -25)/4.9 (42, 4, -30) |
| | Middle Occipital Gyrus | 18, 19 | 0.1/2.1 | 3.9 (-42, -77, 11)/5.8 (39, -76, 1) |
| | Supramarginal Gyrus | 40 | 0.1/0.8 | 3.7 (-46, -42, 37)/5.7 (53, -38, 31) |
| | Medial Frontal Gyrus | 10 | 0.9/2.8 | 5.6 (-12, 57, -4)/5.5 (13, 57, 8) |
| | Superior Frontal Gyrus | 9, 10, 11 | 1.5/1.6 | 5.5 (-13, 53, -11)/5.1 (10, 56, -7) |
| | Superior Parietal Lobule | 5, 7 | 1.0/0.5 | 5.1 (-30, -51, 55)/5.5 (13, -64, 54) |
| | Cuneus | 7, 17, 18, 19 | 1.0/0.6 | 5.5 (-19, -80, 32)/4.6 (25, -87, 11) |
| | Cerebellar Tonsil | * | 0.0/3.1 | -999.0 (0, 0, 0)/5.4 (28, -51, -37) |
| | Middle Frontal Gyrus | 9, 10 | 1.0/1.4 | 4.9 (-31, 49, -5)/5.2 (34, 49, -4) |
| | Sub-Gyral | 40 | 1.9/1.0 | 5.1 (-25, -45, 52)/5.0 (36, -74, -1) |
| | Insula | 13 | 0.0/0.3 | -999.0 (0, 0, 0)/4.9 (58, -33, 18) |
| | Postcentral Gyrus | 2, 3, 40 | 0.2/1.3 | 4.3 (-46, -29, 37)/4.9 (46, -26, 39) |
| | Inferior Temporal Gyrus | 20, 21 | 1.2/0.4 | 4.9 (-49, -1, -28)/4.2 (56, -13, -17) |
| | Superior Temporal Gyrus | 22, 38, 42 | 0.1/1.0 | 3.8 (-49, 6, -22)/4.7 (56, -33, 14) |
| | Fusiform Gyrus | 20 | 0.5/0.2 | 4.7 (-49, -4, -25)/4.3 (52, -3, -25) |
| | * | * | 0.0/0.0 | -999.0 (0, 0, 0)/-999.0 (0, 0, 0) |
| | Angular Gyrus | 39 | 0.6/0.0 | 4.6 (-43, -59, 36)/-999.0 (0, 0, 0) |
| | Cingulate Gyrus | * | 0.3/0.0 | 4.6 (-16, -34, 27)/-999.0 (0, 0, 0) |
| | Precentral Gyrus | 44 | 0.1/0.4 | 3.6 (-49, 18, 7)/4.3 (55, -19, 35) |
| | Paracentral Lobule | 5 | 0.3/0.0 | 4.1 (-19, -41, 50)/-999.0 (0, 0, 0) |
| | Inferior Semi-Lunar Lobule | * | 0.0/0.3 | -999.0 (0, 0, 0)/4.1 (31, -62, -39) |

| Network | Area | Brodmann Area | Volume (cc) | Random effects: Max Value (x, y, z) |
|---|---|---|---|---|
| | Inferior Occipital Gyrus | 19 | 0.0/0.2 | -999.0 (0, 0, 0)/4.0 (36, -77, -4) |
| | Caudate | * | 0.1/0.1 | 3.6 (-9, 17, 12)/4.0 (19, -28, 19) |
| | Inferior Frontal Gyrus | 45 | 0.1/0.0 | 4.0 (-49, 20, 11)/-999.0 (0, 0, 0) |
| | | | | |
| IC10 GM | * | * | 0.1/0.3 | 5.5 (-3, 12, -22)/6.4 (3, 21, -24) |
| | Rectal Gyrus | 11 | 0.4/0.4 | 5.3 (-6, 18, -25)/5.8 (3, 23, -26) |
| | Caudate | * | 1.3/0.3 | 4.7 (-12, 14, 9)/3.9 (15, 15, 8) |
| | Sub-Gyral | * | 0.2/0.3 | 3.6 (-21, 25, 37)/4.4 (42, 19, 23) |
| | Superior Frontal Gyrus | 6, 8, 9 | 1.0/1.2 | 4.3 (-22, 44, 32)/4.3 (10, 10, 58) |
| | Medial Frontal Gyrus | 6 | 0.8/0.4 | 4.1 (-3, -22, 62)/3.9 (1, -11, 50) |
| | Middle Frontal Gyrus | 8 | 0.4/0.2 | 4.1 (-24, 25, 40)/3.7 (45, 22, 22) |
| | Tuber | * | 0.0/0.3 | -999.0 (0, 0, 0)/4.1 (36, -58, -30) |
| | Lentiform Nucleus | * | 0.4/0.6 | 4.0 (-15, 3, -4)/4.0 (18, 3, -3) |
| | Culmen | * | 0.0/0.6 | -999.0 (0, 0, 0)/4.0 (24, -35, -22) |
| | Extra-Nuclear | * | 0.3/0.1 | 4.0 (-15, 11, 9)/3.7 (33, 20, -2) |
| | Inferior Parietal Lobule | * | 0.0/0.1 | -999.0 (0, 0, 0)/4.0 (39, -42, 41) |
| | Insula | 13 | 0.0/0.1 | -999.0 (0, 0, 0)/3.9 (33, 23, 0) |
| | Middle Temporal Gyrus | * | 0.0/0.1 | -999.0 (0, 0, 0)/3.9 (43, -55, 12) |
| | Precuneus | 7 | 0.3/0.0 | 3.8 (-22, -65, 36)/-999.0 (0, 0, 0) |
| | Cerebellar Tonsil | * | 0.0/0.1 | -999.0 (0, 0, 0)/3.7 (33, -58, -32) |
| | Paracentral Lobule | * | 0.1/0.0 | 3.6 (-1, -29, 58)/-999.0 (0, 0, 0) |
| | Inferior Frontal Gyrus | * | 0.1/0.0 | 3.5 (-43, 34, 12)/-999.0 (0, 0, 0) |
| | | | | |
| IC10 WM | Precentral Gyrus | 6, 9, 44 | 1.2/0.9 | 11.0 (-37, 2, 36)/6.6 (45, 5, 36) |

| Network | Area | Brodmann Area | Volume (cc) | Random effects: Max Value (x, y, z) |
|---|---|---|---|---|
| | Sub-Gyral | 31 | 11.9/7.7 | 9.8 (-34, 2, 33)/6.8 (42, 9, 15) |
| | Middle Frontal Gyrus | 6, 8, 9, 10 | 3.0/1.7 | 9.2 (-40, 2, 39)/5.2 (48, 8, 37) |
| | Inferior Frontal Gyrus | 6, 9, 47 | 1.0/0.8 | 9.1 (-37, -1, 33)/5.0 (37, 3, 33) |
| | Precuneus | 7, 19, 31, 39 | 1.0/2.1 | 7.0 (-33, -64, 36)/7.0 (25, -49, 43) |
| | Superior Parietal Lobule | 7 | 0.7/1.0 | 4.8 (-33, -66, 43)/6.9 (28, -50, 45) |
| | Medial Frontal Gyrus | 6, 8, 9, 32 | 3.6/0.6 | 6.5 (-25, 43, 8)/4.2 (22, 43, 17) |
| | Supramarginal Gyrus | * | 1.4/0.1 | 6.5 (-43, -52, 29)/3.5 (46, -52, 34) |
| | Superior Frontal Gyrus | 6, 9, 10 | 2.8/2.7 | 6.1 (-15, 13, 52)/6.4 (21, 55, 15) |
| | Superior Temporal Gyrus | 22, 39 | 1.2/0.1 | 6.2 (-43, -51, 25)/3.6 (55, -37, 17) |
| | Insula | 13 | 0.0/1.7 | -999.0 (0, 0, 0)/6.1 (37, 9, 15) |
| | Inferior Parietal Lobule | 40 | 0.3/1.1 | 5.1 (-30, -53, 47)/5.9 (31, -53, 45) |
| | Cingulate Gyrus | 24, 31, 32 | 3.4/0.4 | 5.9 (-12, 8, 45)/4.7 (19, -43, 37) |
| | Middle Temporal Gyrus | 20, 39 | 1.0/0.7 | 5.8 (-33, -63, 27)/5.0 (55, -33, -12) |
| | Anterior Cingulate | 9, 32 | 0.0/0.8 | -999.0 (0, 0, 0)/5.7 (15, 36, 12) |
| | Cerebellar Tonsil | * | 1.8/0.8 | 5.7 (-25, -45, -32)/5.1 (22, -45, -32) |
| | * | * | 0.4/0.4 | 4.8 (-22, -42, -29)/4.5 (21, -42, -29) |
| | Cuneus | 17, 18, 30 | 2.4/1.8 | 5.2 (-9, -76, 23)/5.5 (19, -73, 6) |
| | Middle Occipital Gyrus | 18 | 1.2/0.4 | 5.4 (-16, -90, 14)/3.9 (16, -88, 14) |
| | Lingual Gyrus | 18 | 0.1/1.4 | 4.1 (-31, -72, -4)/5.0 (10, -85, -3) |
| | Paracentral Lobule | 5 | 0.5/0.1 | 4.9 (-16, -35, 50)/3.7 (18, -42, 57) |
| | Angular Gyrus | * | 0.3/0.0 | 4.8 (-34, -58, 35)/-999.0 (0, 0, 0) |
| | Posterior Cingulate | 30, 31 | 0.6/0.1 | 4.7 (-13, -65, 14)/4.3 (25, -67, 6) |

| Network | Area | Brodmann Area | Volume (cc) | Random effects: Max Value (x, y, z) |
|---|---|---|---|---|
| | Fusiform Gyrus | 20, 37 | 0.3/0.1 | 4.6 (-30, -53, -9)/3.5 (49, -36, -21) |
| | Inferior Occipital Gyrus | * | 0.1/0.0 | 4.5 (-34, -74, -4)/-999.0 (0, 0, 0) |
| | Extra-Nuclear | * | 0.3/0.6 | 3.8 (-22, 18, -10)/4.5 (33, 10, 16) |
| | Postcentral Gyrus | 3 | 0.1/0.3 | 3.5 (-30, -21, 43)/4.1 (18, -38, 59) |
| | Lateral Ventricle | * | 0.2/0.3 | 4.1 (-4, 19, 0)/3.8 (28, -64, 6) |
| | Rectal Gyrus | 11 | 0.0/0.1 | -999.0 (0, 0, 0)/4.0 (6, 21, -25) |
| | Inferior Temporal Gyrus | 20 | 0.1/0.3 | 3.6 (-39, -70, 0)/3.9 (52, -33, -15) |
| | Declive of Vermis | * | 0.1/0.0 | 3.9 (-1, -69, -15)/-999.0 (0, 0, 0) |
| | Culmen | * | 0.1/0.0 | 3.8 (-33, -44, -29)/-999.0 (0, 0, 0) |
| | Caudate | * | 0.1/0.0 | 3.6 (-7, 19, -2)/-999.0 (0, 0, 0) |

**Table 3**

**Result from independent sample t-Test analysis**

| | Group | N | Mean | SD | |
|---|---|---|---|---|---|
| Age (Ave) | Low Accepters | 70 | 29.929 | 12.71 | t (126) = 0.201, p = 0.580, d= -0.036 |
| | High Accepters | 58 | 29.483 | 12.21 | |
| gender (F=1, M=2) | Low Accepters | 70 | F=20 | 0.455 | t (126) = -0.122, p = 0.451, d= -0.022 |
| | High Accepters | 58 | F=16 | 0.451 | |
| NEOFFI_ Neuroticism | Low Accepters | 70 | 1,40 | 0,60 | t(124.44) = -1.311, p = 0.096, d = 0.232 |
| | High Accepters | 58 | 1,54 | 0,55 | |
| NEOFFI_ Extraversion | Low Accepters | 70 | 2,56 | 0,53 | t(118.57) = 1.403, p = 0.918, d = 0.250 |
| | High Accepters | 58 | 2,40 | 0,51 | |
| NEOFFI_ Openness to Experiences | Low Accepters | 70 | 2,61 | 0,52 | t(122.92) = -1.760, p = 0.040*, d = -0.312 |
| | High Accepters | 58 | 2,76 | 0,50 | |

| | Group | N | Mean | SD | |
|---|---|---|---|---|---|
| NEOFFI_ Agreeableness | Low Accepters | 70 | 2,77 | 0,41 | t(116.10) = -1.152, p = 0.874, d = -0.206 |
| | High Accepters | 58 | 2,80 | 0,45 | |
| NEOFFI_ Conscientiousness | Low Accepters | 70 | 2,74 | 0,54 | t(114.69) = 1.623, p = 0.946, d = 0.290 |
| | High Accepters | 58 | 2,53 | 0,63 | |
| TeiQueSF self_control | Low Accepters | 70 | 5.096 | 0.811 | t(124.48) = 0.191, p = 0.576, d = 0.034 |
| | High Accepters | 58 | 5.069 | 0.749 | |
| TeiQueSF emotionality | Low Accepters | 70 | 5.17 | 0.765 | t(114.68) = 1.345, p = 0.909, d = 0.240 |
| | High Accepters | 58 | 4.974 | 0.869 | |
| TeiQueSF sociability | Low Accepters | 70 | 5.111 | 0.635 | t(97.59) = 1.970, p =0.974, d = 0.356 |
| | High Accepters | 58 | 4.828 | 0.93 | |
| TeiQueSF well_being | Low Accepters | 70 | 5.793 | 0.805 | t(115.73) = 0.824, p =0.794 , d = 0.147 |
| | High Accepters | 58 | 5.667 | 0.899 | |
| TeiQueSF_total | Low Accepters | 70 | 156 | 15 | t(106.61) = 1.437, p =0.923, d = 0.258 |
| | High Accepters | 58 | 151,81 | 19,58 | |

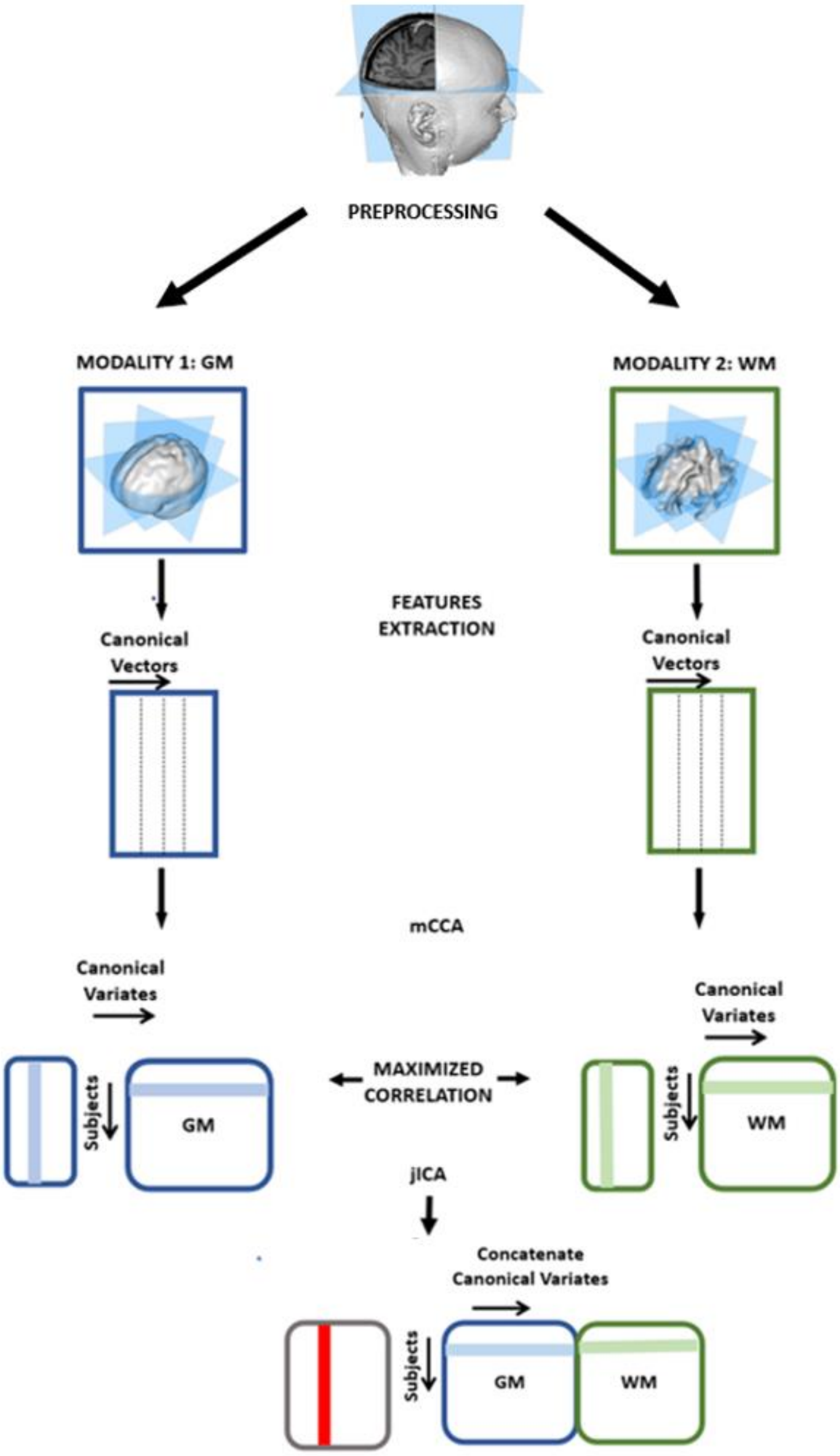

**Fig. 1**

mCCA+jICA method: First, the structural T1 images were preprocessed then the GM and WM features were extracted. Next, the features were reorganized into two matrices and the dimensionality was reduced. Furthermore, mCCA was applied to the feature matrices then canonical variants and the associated components were computed. Finally, jICA was applied to compute the maximized joint-independent component matrices and the relative loading coefficients mixing matrix.

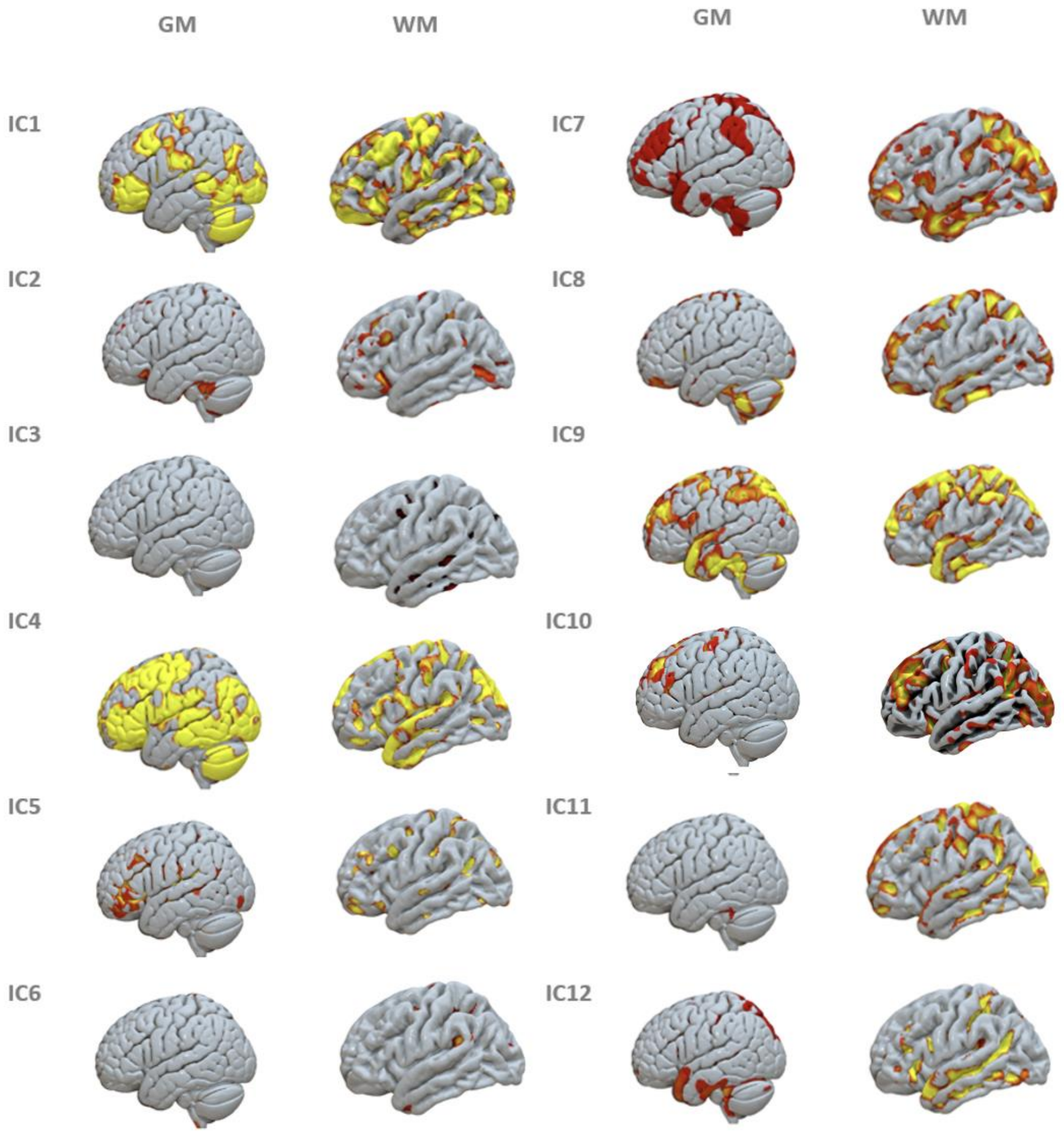

**Fig. 2:** Independent components from 1 to 12. ICA was able to decompose the brain into 12 covarying gray and white matter networks

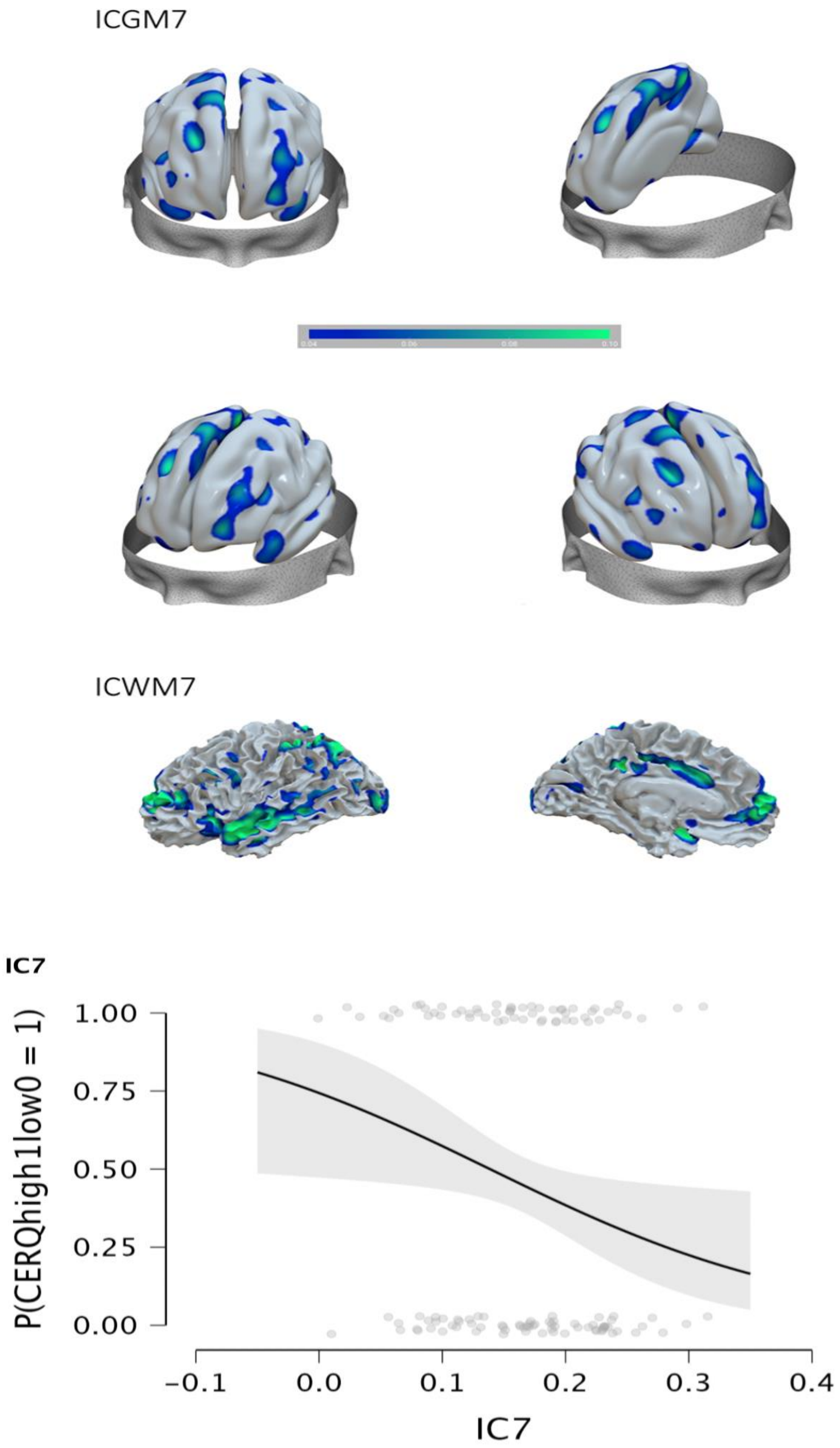

**Fig. 3:** Brain plots of Network 1. Top. Brain plot reconstruction of IC7 gray matter concentration of high accepters. Central: gray and white matter concentration of IC7. Bottom: loading coefficients distribution between low and high accepters.

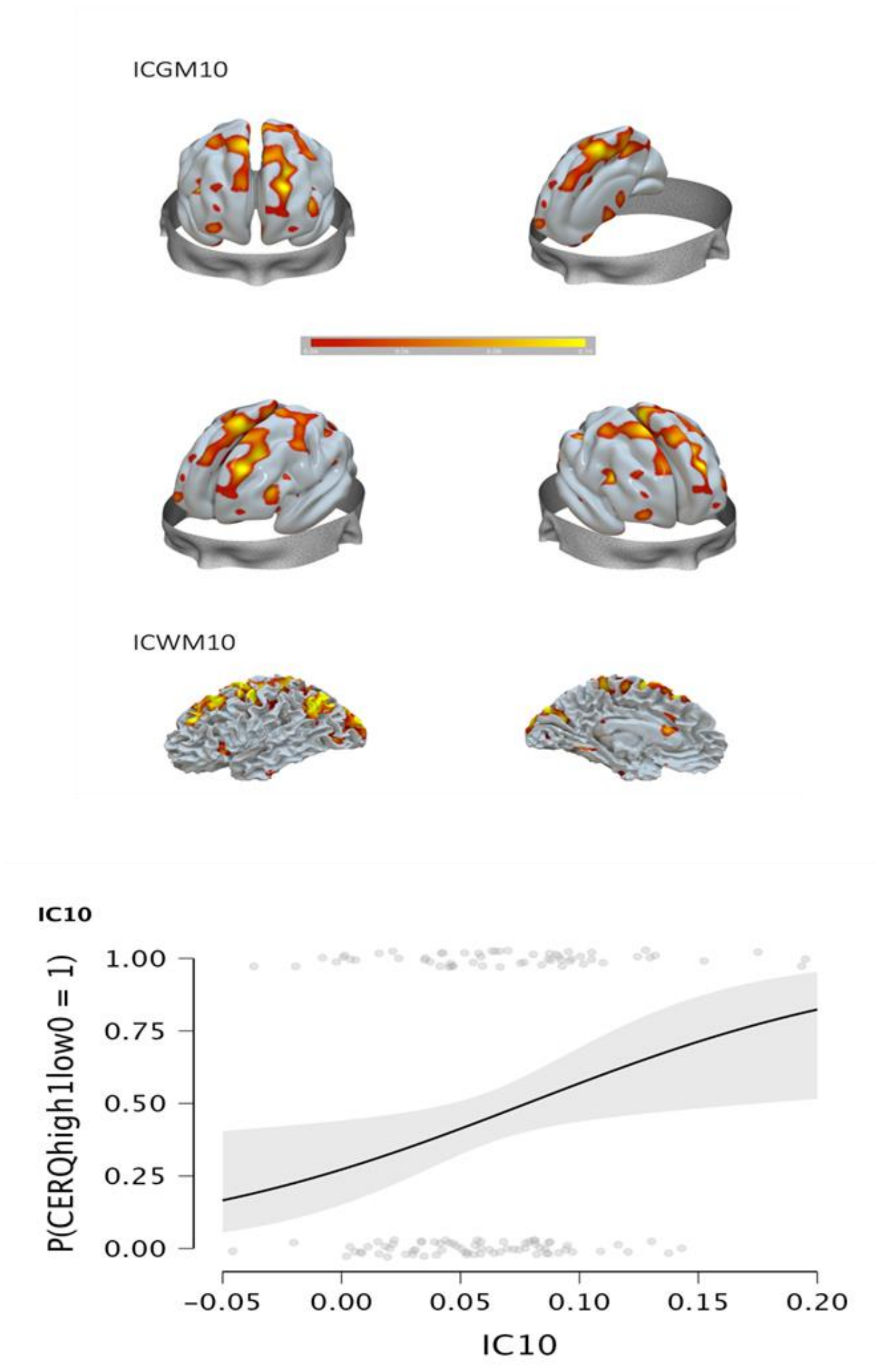

**Fig. 4:** Brain plots of Network 2. Top. Brain plot reconstruction of IC10 gray matter concentration of high accepters. Central. White matter concentration of IC10. Bottom: loading coefficients distribution between low and high accepters.

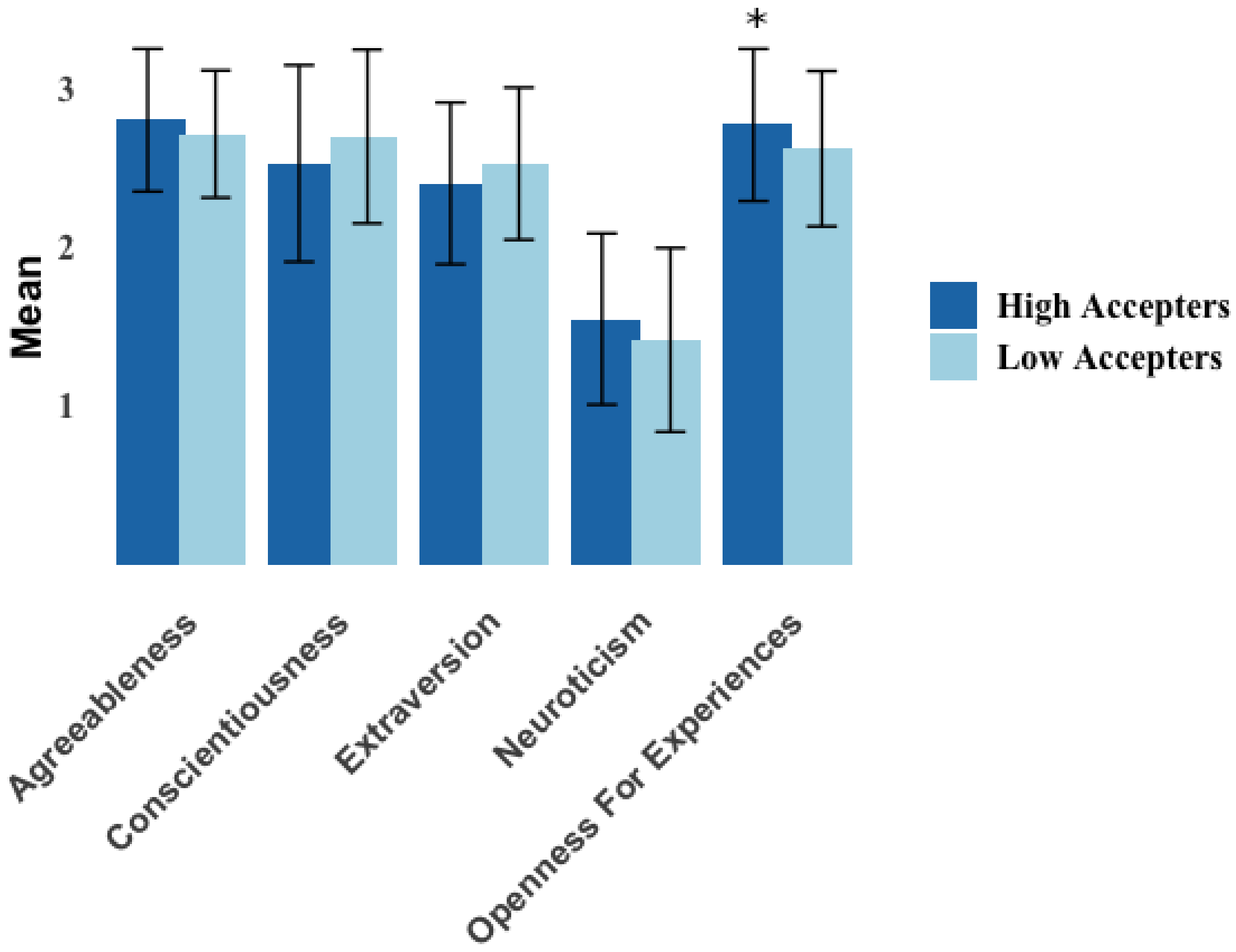

*Note:* * for p < .05,

**Fig 5**: Personality subscales for low vs high accepters

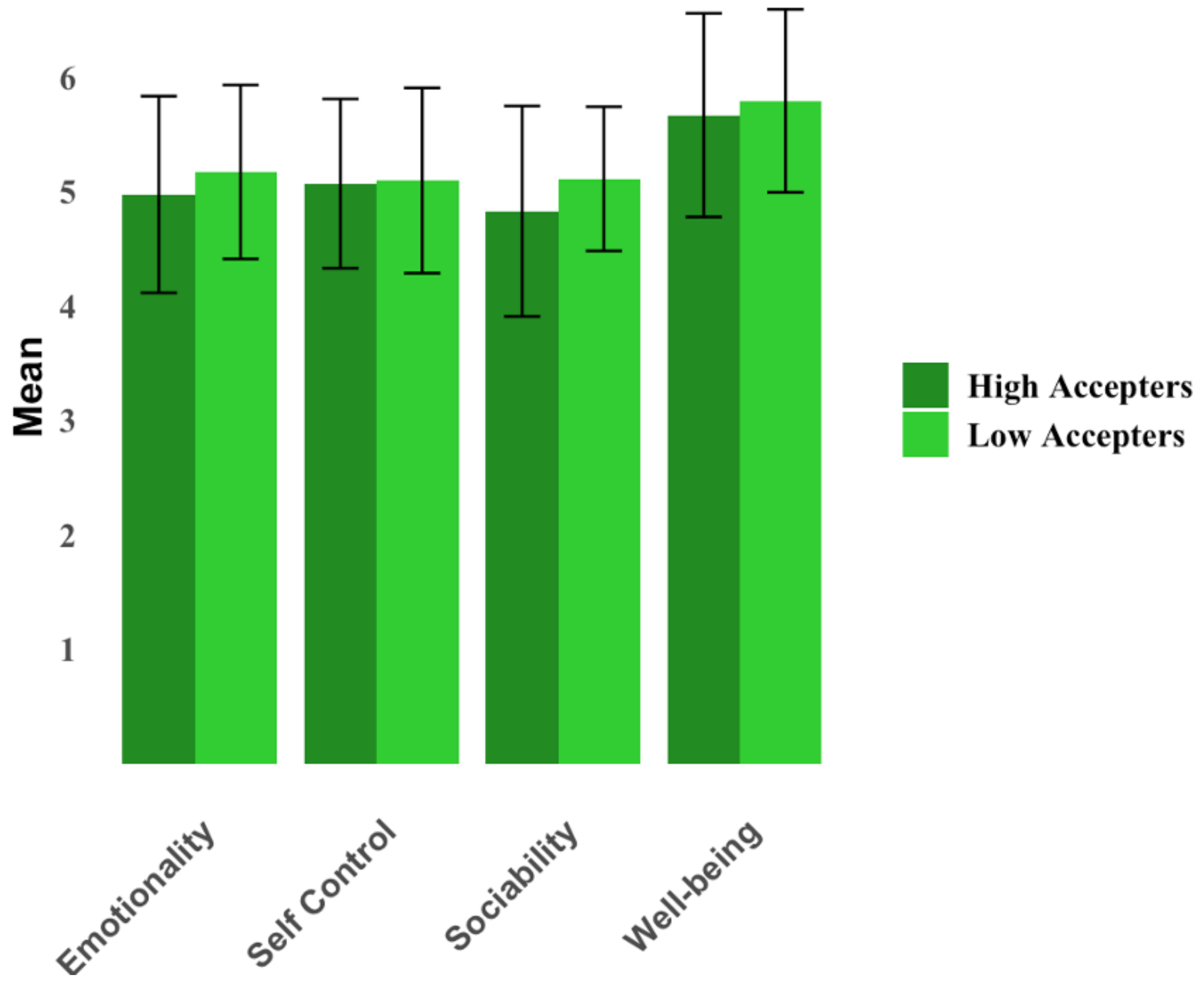

**Fig 6:** Emotional intelligence subscales for low vs high accepters